# An 11.7-GHz ScAlN FBAR Filter: Case Study on Scaling Limits and Challenges

Sinwoo Cho*, Byeongjin Kim*, Lezli Matto, Omar Barrera, Pietro Simeoni, Yinan Wang, Michael Liao, Tzu-Hsuan Hsu, Jack Kramer, Matteo Rinaldi, Mark S. Goorsky, and Ruochen Lu

***Abstract*— This paper reports an 11.7-GHz compact 50-Ω ladder filter based on single-layer Scandium Aluminum Nitride (ScAlN) film bulk acoustic resonators (FBARs) with platinum (Pt) electrodes, and uses it as a quantitative case study of the limits encountered when directly scaling to higher frequencies. The measured filter achieves a 3-dB fractional bandwidth (FBW) of 4.0% and an out-of-band rejection greater than 23.1 dB, with a minimum insertion loss (IL) of 6.8 dB. We analyze the origin of this performance through a quantitative framework: (i) a loss decomposition study, (ii) frequency-shift sensitivity that explains the discrepancy between simulated and measured center frequency, (iii) FBW sensitivity to series–shunt separation and port impedance, and (iv) stress-limited aperture that constrains device size. The results establish a realistic, fabricable baseline for directly scaled single-layer ScAlN FBAR filters and outline materials, electrode, and stress-management directions toward lower-loss mmWave acoustic filters.**

## I. INTRODUCTION

THE relentless expansion of wireless communication into 5G and future 6G systems is driving a persistent demand for components that can operate at higher frequencies and support wider bandwidths [1], [2]. As the usable spectrum below 6 GHz becomes increasingly congested, the industry is migrating to centimeter-wave (e.g., Ku band) and millimeter-wave frequencies to unlock the vast bandwidth required for emerging applications [3]–[5]. This migration places extreme demands on RF front-end components, particularly filters, which must provide sharp frequency selectivity in a compact form factor [6], [7].

Acoustic wave technology has become the cornerstone of RF filtering in mobile devices due to its profound miniaturization advantage [8], [9]. By converting electromagnetic signals into mechanical vibrations, acoustic devices operate at wavelengths that are four to five orders of magnitude shorter than their electromagnetic counterparts, enabling filter sizes that are thousands of times smaller [10], [11]. Among acoustic technologies, Film Bulk Acoustic Resonators (FBARs) are particularly well-suited for high-frequency operation [12], [13]. In an FBAR, the acoustic energy is confined within the bulk of a thin piezoelectric film, leading to high Quality ($Q$) factors compared to surface acoustic wave (SAW) devices [14], [15].

The material platform for FBARs has evolved to meet the demands for wider bandwidth [16], [17]. Scandium-doped Aluminum Nitride (ScAlN) has emerged as a leading material, offering a significantly higher electromechanical coupling coefficient ($k^2$) than traditional AlN [18], [19]. This enhanced coupling is essential for designing the wide-bandwidth filters stipulated by modern communication standards [20], [21].

However, the path to higher frequencies is not without significant obstacles [22]. The conventional method for increasing an FBAR's operating frequency is to reduce the thickness of the piezoelectric and metal layers, as the fundamental thickness extensional resonant frequency is inversely proportional to this critical dimension [23]–[25]. While straightforward in principle, this scaling approach introduces a cascade of practical challenges that degrade the resonator's $Q$ factor . Since the insertion loss (IL) of a filter is inversely related to the $Q$ of its constituent resonators, this degradation creates a major bottleneck for achieving the low-loss RF front ends [26], [27].

In this work, we report an 11.7-GHz, 50-Ω ladder filter using a single-layer $Sc_{0.3}Al_{0.7}N$ FBAR with Pt electrodes and utilize it as a quantitative case study of direct thickness scaling. The measured device achieves FBW of 4.0%, out-of-band (OoB) rejection of 23.1 dB, and minimum IL of 6.8 dB at 11.7 GHz. Rather than a performance milestone, our contribution is primarily a baseline and loss decomposition, which involves quantifying how various factors, such as thin-film crystal quality, electrode series resistance, and residual film stress, map onto IL, FBW, and center-frequency shift. We further contextualize these results against representative state-of-the-art (SoA) filters above 10 GHz [28], [29], e.g., periodically poled piezoelectric film (P3F) ScAlN and thin-film lithium niobate (TFLN) laterally field excited bulk acoustic resonator (XBAR) filters, which typically have lower IL and wider FBW at the cost of greater stack complexity, to calibrate expectations for single-layer ScAlN implementations. The paper first presents the device, fabrication, and measured results, then develops a quantitative loss budget and sensitivity analysis, and finally outlines materials, electrodes, and stress-management directions to reduce loss in future mmWave filters.

Manuscript received XX 2025; revised XX June 2025; accepted XX June 2025. This work was supported by DARPA COmpact Front-end Filters at the ElEment-level (COFFEE).

S. Cho, B. Kim, O. Barrera, Y. Wang, T.-H. Hsu, J. Kramer, and R. Lu are with The University of Texas at Austin, Austin, TX, USA (email: sinwoocho@utexas.edu). P. Simeoni and M. Rinaldi are with Northeastern University, Boston, MA, USA. L. Matto, M. Liao, and M. S. Goorsky are with University of California, Los Angeles, Los Angeles, CA, USA.

S. Cho and B. Kim contribute equally to the work.

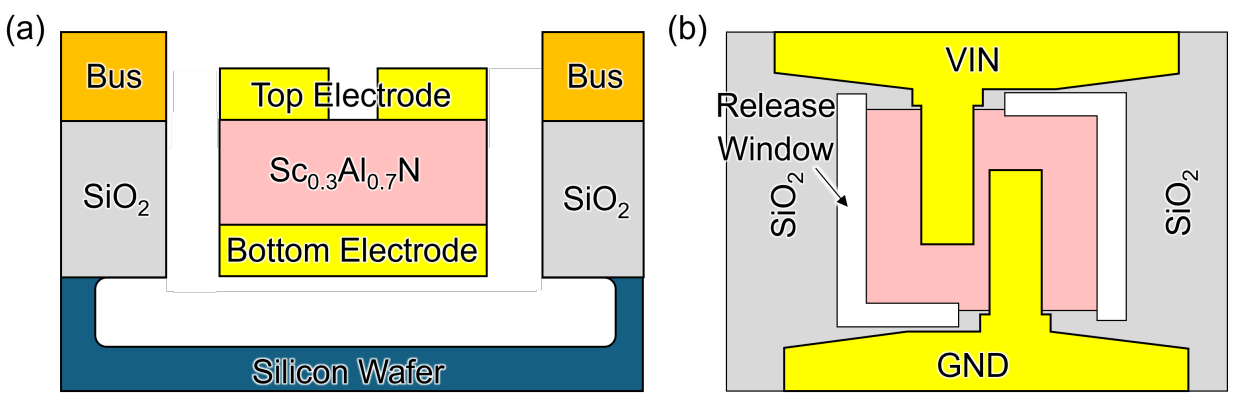

Fig. 1 (a) Cross-sectional and (b) top view of thin-film ScAlN FBAR.

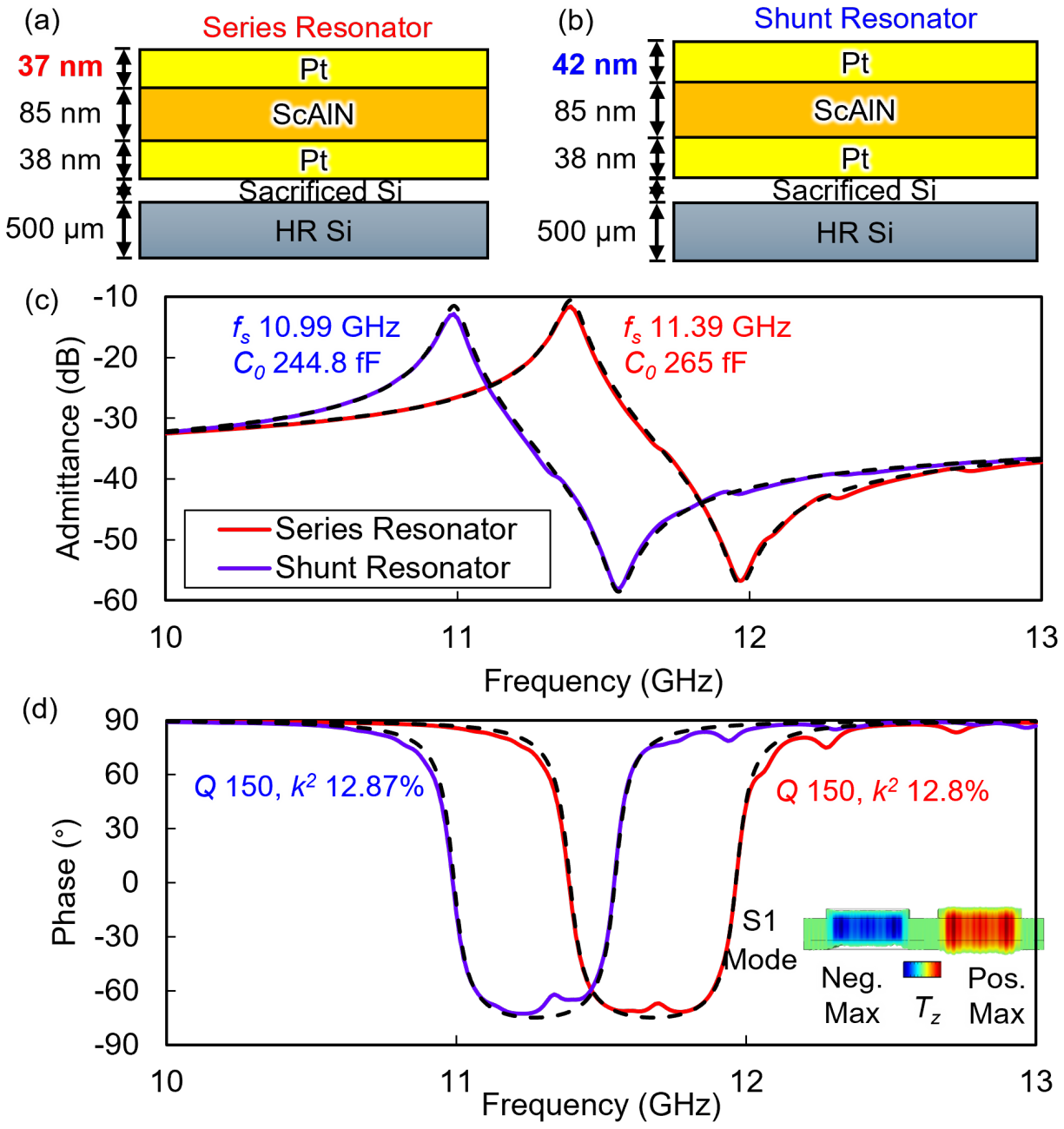

Fig. 2 (a) Series and (b) shunt resonator with different top Pt electrode thickness. Simulated wideband admittance (c) amplitude and (d) phase with simulated displacement mode shape for series and shunt resonator, respectively, where the dotted lines represent the BVD fitting.

## II. FILTER TESTBED DESIGN AND BASELINE PERFORMANCE

To quantify the limits of direct thickness scaling in single-layer ScAlN FBARs, we implement a ladder-filter testbed using dual top Pt electrodes to achieve mass-loaded detuning of resonance. A 7th-order (4 series/3 shunt) topology is synthesized and fabricated on high-resistivity Si. We first establish the measured baseline at 11.7 GHz. These simulations and measurements set the reference point for the quantitative loss decomposition and scaling-challenge analysis that follows.

### *A. Architecture and Design Rationale*

The FBAR prototype consists of two top electrodes, connecting to VIN (drive) and GND (ground), respectively, a piezoelectric layer, and a floating bottom electrode, effectively creating two resonators in series. This design is selected because it requires a straightforward fabrication process that does not require patterning the bottom electrode prior to ScAlN sputtering. To implement the filter, the series and shunt resonators were based on an 85 nm thick $Sc_{0.3}Al_{0.7}N$ piezoelectric layer with a 38 nm Pt bottom electrode, as shown in Fig. 1. More details of the resonator design and comparison of different bottom electrode metals have been reported in our prior works in [30].

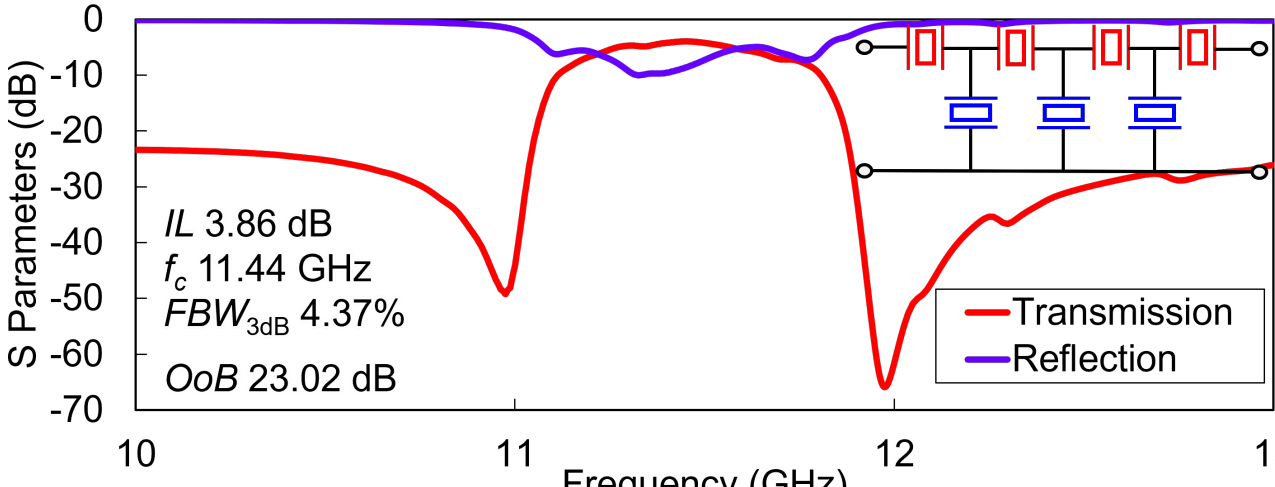

Fig. 3 Ladder structure filter, and simulated filter response.

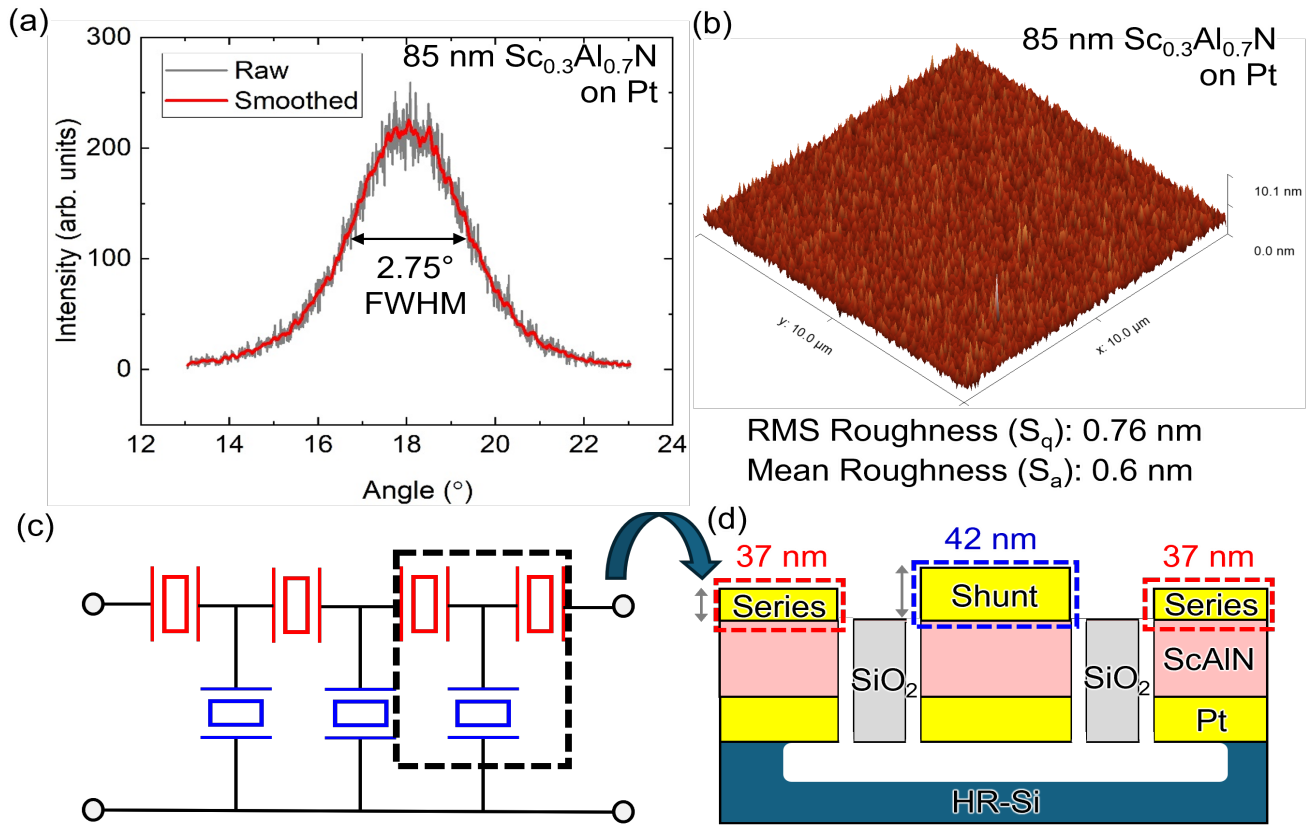

Fig. 4 (a) XRD symmetric rocking curve of 85 nm sputtered thin-film ScAlN. (b) ScAlN film surface roughness measurement by AFM [30]. (c) Ladder structure filter and (d) cross-sectional view of the fabricated ScAlN BAW filter.

The filter was designed using a ladder topology, which requires series and shunt resonators with shifted resonant frequencies. This frequency detuning was achieved by mass-loading, where the top electrode of the shunt resonators is made slightly thicker than that of the series resonators. The frequency separation was achieved by designing the series resonators with a 37 nm-thick Pt top electrode and the shunt resonators with a thicker 42 nm Pt top electrode, as shown in Fig. 2. COMSOL Finite Element Analysis (FEA) simulations were conducted to optimize these thicknesses and predict the resonators' performance. The simulations indicated that resonators with a $k^2$ of 12.8% were achievable (Fig. 2). Note that such high $k^2$ tends to be hard to achieve, due to parasitic introduced layout and fabrication, as well as limited film quality. The key parameters are extracted from a BVD fitting (dotted lines) and listed in Fig. 2 (c)(d). A $Q$ factor of 150 is used here, based on prior measurements of similar devices. The impact of $Q$ will be discussed in later discussion sections. The ScAlN material constants used in the FEA were reported in [31]. However, it is notable that the material parameters, such as density, stiffness, dielectric, and piezoelectric constants, are likely to vary in thinner films as the crystal quality worsens. Future dedicated study on the impact of material parameters in ScAlN films of different thicknesses, crystal qualities, and substrates would be of crucial impact in the research area.

The primary purpose of this work is to synthesize resonators into filters. The resonators simulated above are used to synthesize a filter in Figure 6. Using the designed resonators, a full 7th-order (4 series, 3 shunt) ladder filter was simulated for a 50 Ω system impedance. The simulation results predicted a

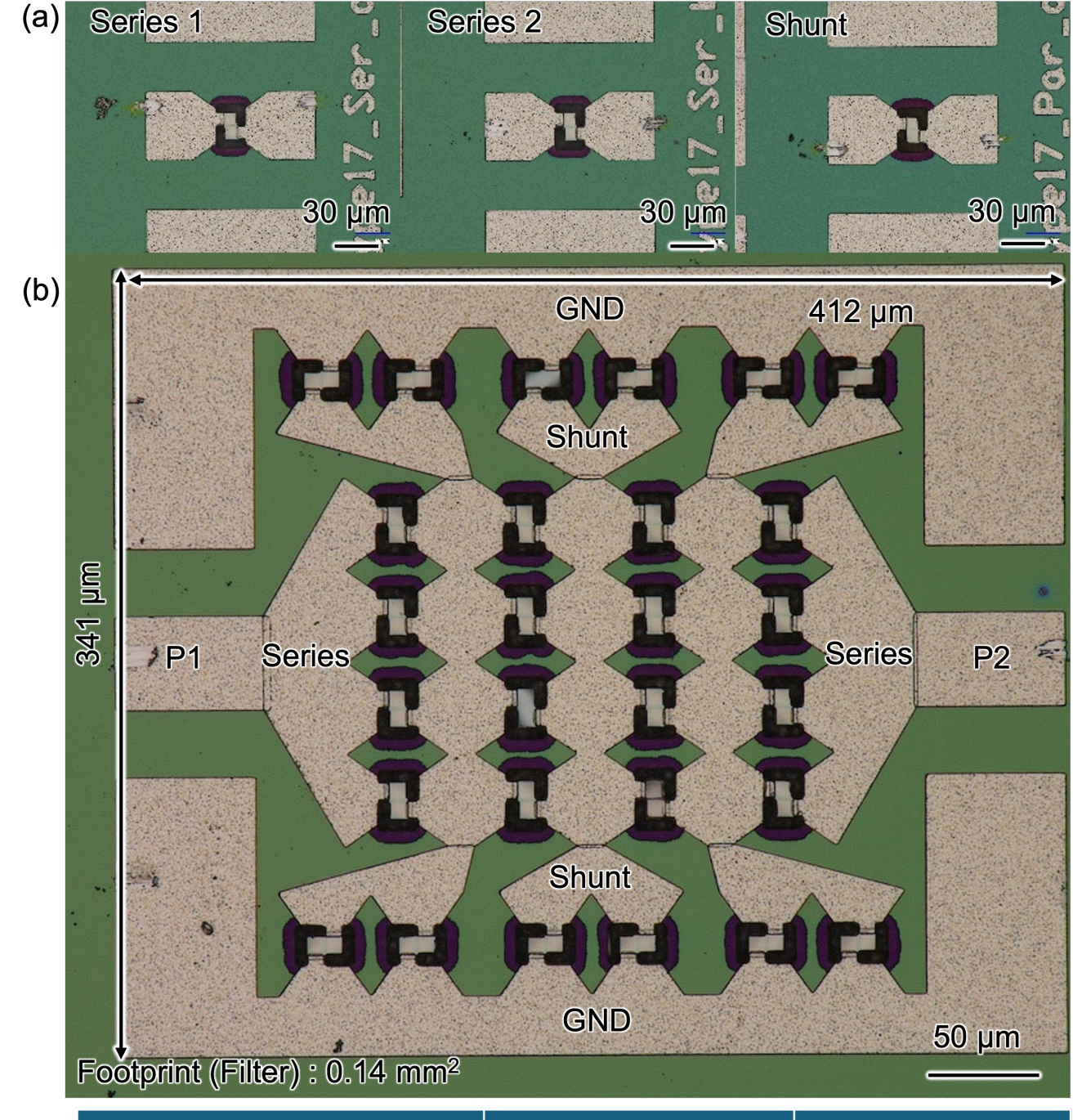

| Parameter | Series 1&2 | Shunt |
|---|---|---|
| Top electrode thick. (nm) | 34.5 | 38.9 |
| ScAlN thickness (nm) | 85 | 85 |
| Active region dim. (μm) | 17.1x9.0, 17.8x9.0 | 16.8x9.0 |
| Top electrode dim. (μm) | 6.3x8.5, 6.6x8.5 | 6.1x8.5 |
| Dis. bet. electrodes (μm) | 1.75, 1.80 | 1.80 |
| Busline thick. (nm) | 350 | 350 |

Fig. 5 (a) Microscopic images of fabricated series resonators, shunt resonator, and (b) filter. (b) Key dimensions are listed in the table.

clear bandpass response at the center frequency $f_c$ of 11.44 GHz, with IL of 3.86 dB, FBW of 4.37%, and a high out-of-band (OoB) rejection of 23.02 dB. The return loss could be improved with future optimization. These simulations confirmed the design's viability and provided a performance baseline for the fabricated device.

## *B. Fabrication Overview*

The fabrication process is illustrated in Fig. 4. The filter was fabricated on a high-resistivity silicon (HR-Si) wafer to minimize substrate-related RF losses. The fabrication process starts with sputter deposition of two layers: first, a 38 nm Pt film, and then an 85 nm ScAlN film . An Evatec Clusterline 200 is used to perform the deposition under an uninterrupted vacuum. The quantitative material analysis begins with X-ray diffraction (XRD), as shown in Fig. 4(a). The FWHM of the rocking curve is 2.75°, indicating that the sputtered thin film has good crystal quality, given that it is sputtered on top of metal with an overall thickness of less than 100 nm. The atomic force microscopy (AFM) in Fig. 4 (b) shows the RMS surface roughness of the sputtered ScAlN film with 0.76 nm. The surface is generally flat and has good uniformity [30]. After that, the regions outside the active resonator areas, composed of ScAlN, Al, Si, and bottom Pt layers, are etched by AJA Ion Mill [32]. The etched regions are then passivated with a low-temperature (100 ◦C) plasma-enhanced chemical vapor deposition (PECVD) process, resulting in the deposition of 360 nm of SiO2, which provides electrical isolation while preventing top electrode disconnection due to the height steps. Next, the top electrodes are deposited and patterned, using 37 nm Pt for series resonators and 42 nm Pt for shunt resonators. Afterward, a 300 nm Al layer is deposited to form the buslines and contact pads. This is followed by the etch window definition, which exposes the silicon substrate beneath the active resonator areas. Finally, acoustic isolation is achieved using $XeF_2$ gas-phase etch to isotropically remove silicon beneath the resonators, resulting in suspended membrane structures that acoustically isolate the devices from the substrate and thus help ensure high-$Q$ performance.

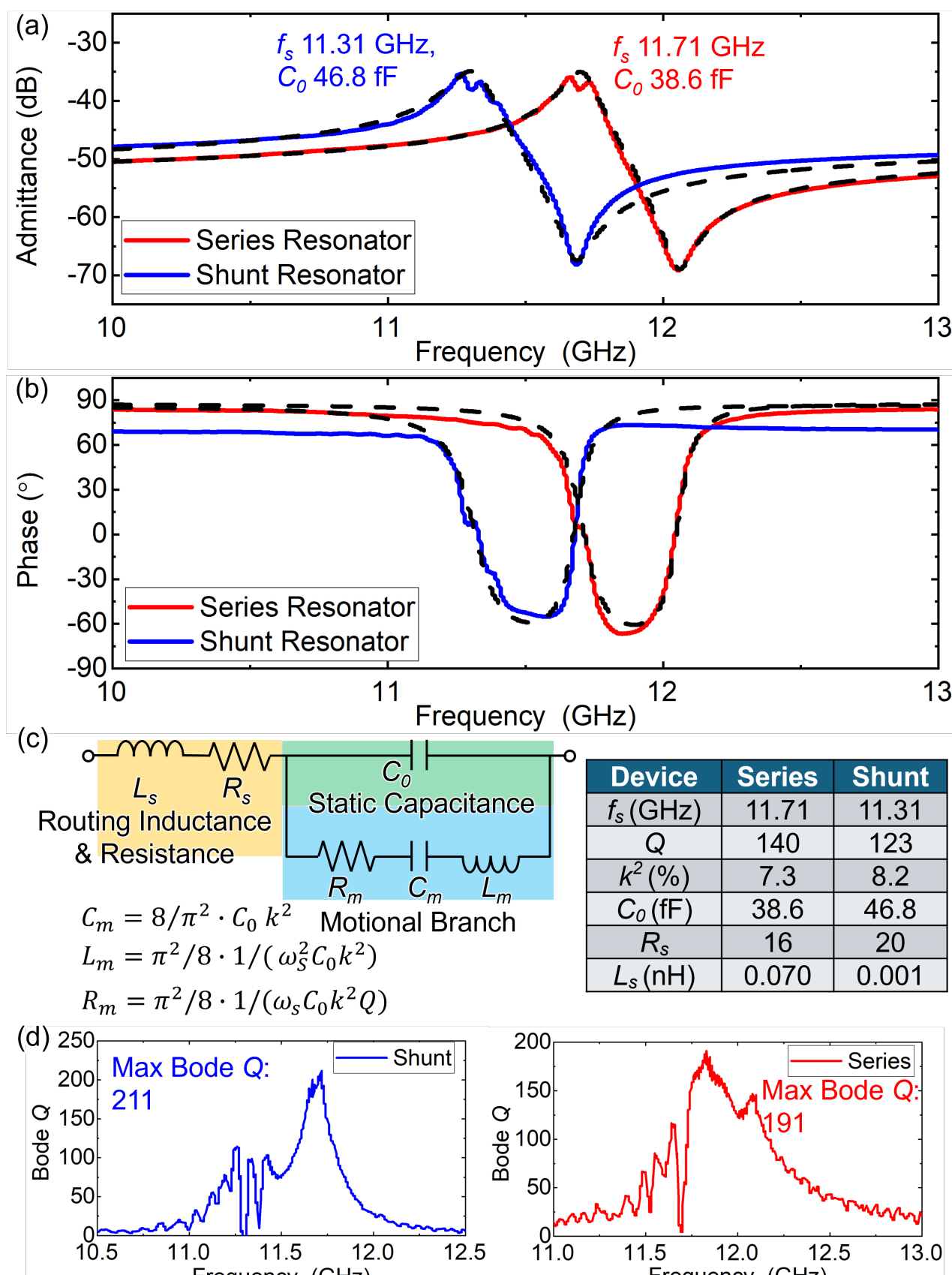

| Device | Series | Shunt |
|---|---|---|
| $f_s$ (GHz) | 11.71 | 11.31 |
| $Q$ | 140 | 123 |
| $k^2$ (%) | 7.3 | 8.2 |
| $C_0$ (fF) | 38.6 | 46.8 |
| $R_s$ | 16 | 20 |
| $L_s$ (nH) | 0.070 | 0.001 |

Fig. 6 Measured wideband admittance response. (a) Amplitude and (b) phase. (c) Modified mmWave MBVD model and extracted key resonator specifications. (d) Bode $Q$ for series and shunt resonator.

The fabricated series/shunt resonators, as well as the filters, are shown in Fig. 5 (a)-(b), with the key dimensions listed in the inset table. The resonators are split into smaller devices to avoid the stress issue mentioned above. The fabricated filter has a miniature footprint of 0.14 mm², while the device footprint is 0.004 mm², with the remaining area allocated for routing. This design could be further optimized to reduce the footprint.

## *C. Measurement Results and Parameter Extraction*

The fabricated filters and individual test resonators were on-wafer characterized using a vector network analyzer (VNA) at room temperature with a power of -15 dBm. Individual series and shunt test resonators were measured to extract their fundamental parameters, as shown in Fig. 6(a)-(b). By fitting the measured admittance data to a modified BVD equivalent circuit model (Fig. 6c), the following parameters were extracted, as shown in the inset table of Fig. 6.

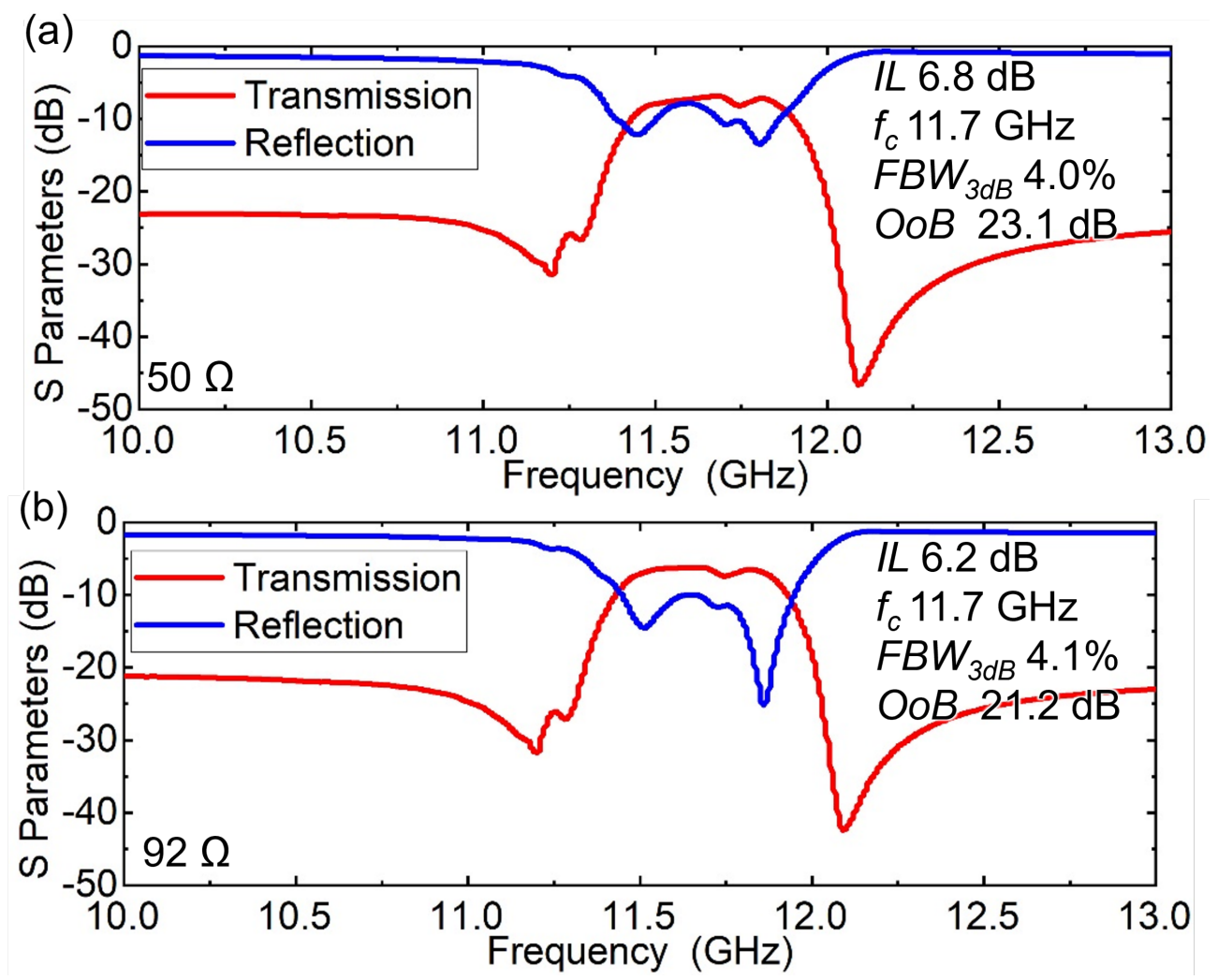

Fig. 7 Measured filter wideband transmission and reflection with (a) 50 Ω and (b) matched to 92 Ω port impedance.

Table I Comparison to State of the Art Filters above 10 GHz

| Reference | $f_c$ (GHz) | IL (dB) | FBW (%) | OoB (dB) | Footprint ($mm^2$) | Technology |
|---|---|---|---|---|---|---|
| [78] | 22.1 | 1.6 | 19.8 | 12.5 | 0.56 | Single LN |
| [79] | 23.8 | 1.5 | 19.4 | 12.1 | 0.64 | P3F LN |
| [80] | 17.4 | 3.3 | 3.4 | 16.6 | 0.28 | P3F ScAlN |
| [81] | 11.9 | 1.5 | 6.6 | 26.0 | 0.25 | P3F ScAlN |
| [82] | 23.8 | 3.8 | 3.4 | - | - | Air-gap AlN |
| [83] | 15 | 3.5 | 10 | 18 | 0.24 | Single ScAlN |
| [84] | 20.9 | 3.2 | 10.96 | 17 | 0.06 | Single ScAlN |
| **This work** | **11.7** | **6.8** | **4.0** | **23.1** | **0.14** | **Single ScAlN** |

The measured resonators exhibited the desired frequency separation between series and shunt resonators (11.31 GHz and 11.71 GHz), and showed $Q$ (parallel resonance $Q$) of 140 and 123, and $k^2$ of 7.3% and 8.2% values for series and shunt resonators, respectively, comparable with prior works using similar stacks in [33]. To further validate, the Bode $Q_s$ of the series and shunt resonators were investigated, and exhibit maximum Bode Q values of 211 and 191, respectively.

The comparison between simulations and measurements, showing lower measured $C_0$ (244.8 to 187.2 fF shunt; 265 to 154.4 fF series) and reduced $k^2$ (12.8% to 8.2%/7.3%) relative to simulation, requires introducing a parasitic series capacitor, $C_s$ (58 fF shunt; 110 fF series), to reconcile both, consistent with our earlier LN results [34] and indicating that the Pt-Al busline probing-pad interface, rather than the resonator body, is likely the dominant source of the discrepancy. Despite the $C_0$ and $k^2$ deviations observed in isolated resonator measurements, the measured filter response maintains better matches with the simulations than the resonators, indicating that pad-induced parasitics in the test fixture have less impact at the filter level.

In addition, the shunt resonator exhibits an out-of-band phase deviation from the capacitive value, likely due to a contacting issue between the busline and electrode metal layers. The issues cause the shunt resonator to behave experimentally with a smaller capacitance than the series one, deviating from the design. However, we expect the issues not to exist in the shunt resonators of the final filter, as the rejection band shows good isolation.

Additionally, one key parameter shown in Fig. 6(c) is the series resistance, $R_s$, which is around 20 Ω for both series and shunt resonators. Note that the fitting here overestimates the actual $R_s$, as the probing pads and routing introduce additional resistive loss. Nevertheless, it indicates that a significant amount of energy is lost as Joule heating. $R_s$ will be discussed in Section II for consideration of the filter response. Self-inductance, $L_s$, is less of a concern here than in other technologies, such as XBARs, because the device is compact and has a high capacitance density.

The S-parameters of the complete ladder filter are shown in Fig. 7. The device showed a clear passband centered at 11.7 GHz. The key measured performance metrics are fc of 11.7 GHz, a minimum IL of 6.8 dB, a 3 dB FBW of 4.0%, a footprint of 0.14 $mm^2$, and an OoB rejection of 23.1 dB. Compared to the simulation, the measured filter had a slightly narrower bandwidth than predicted by the initial simulation, likely due to smaller frequency shifts during the final control of metal loading in the series/shunt resonators. The dip in return loss (RL) around the center frequency is another indicator. The center frequency was slightly higher than the designed value, also due to the metal loading control step. The insertion loss, while higher than the ideal simulation, is mostly due to the routing resistance and the series resistance of the metal, which will be discussed in the next Section. Another contributor is the impedance mismatch due to the frequency setting mentioned above, as the filter shows a lower IL of 6.2 dB when matched to a 92 Ω port impedance, as shown in Fig. 7(b).

The performance validates the design principle and showcases the achievable results with single-layer ScAlN-based FBAR filters. The achieved results are compared with other SoA filter works above 10 GHz in Table I, in terms of important filter performance factors, e.g., $f_s$, IL, FBW, OoB, and footprint. Compared with P3F ScAlN and TFLN XBAR filters, which typically achieve lower IL and wider FBW at similar frequencies through higher-$Q$ materials and engineered multilayer stacks, the directly scaled single-layer ScAlN FBAR reported here incurs higher IL and modest FBW but uses a substantially simpler process/stack, providing a fabricable baseline for isolating loss mechanisms.

## III. Quantitative Analysis of Scaling Challenges

The measured filter validates the ladder topology but reveals non-idealities relative to the pre-fabrication baseline, $f_c$ shifted from 11.45 GHz (sim.) to 11.7 GHz (meas.), IL increased from 3.86 dB to 6.8 dB, and FBW slightly reduced from 4.37% to 4.0%, with partial recovery to 6.2 dB when re-matched to 92 Ω, indicating a measurable mismatch component. The aforementioned challenges of implementing high-performance FBAR are indicative of systemic challenges in scaling FBAR technology [35]. To be specific, to push FBARs to higher frequencies, the piezoelectric ScAlN layer as well as the top and bottom electrodes must be scaled to thicknesses of a few tens of nanometers, causing a few issues: degradation of piezoelectric thin-film crystal quality, increased series resistance in ultra-thin electrodes, challenges in accurately control the film stack thickness during fabrication, and residual film stress that limits device size and structural integrity. We now quantify the origins and impacts of these deviations. The following section used Keysight Advanced Design System (ADS) for the circuit simulation.

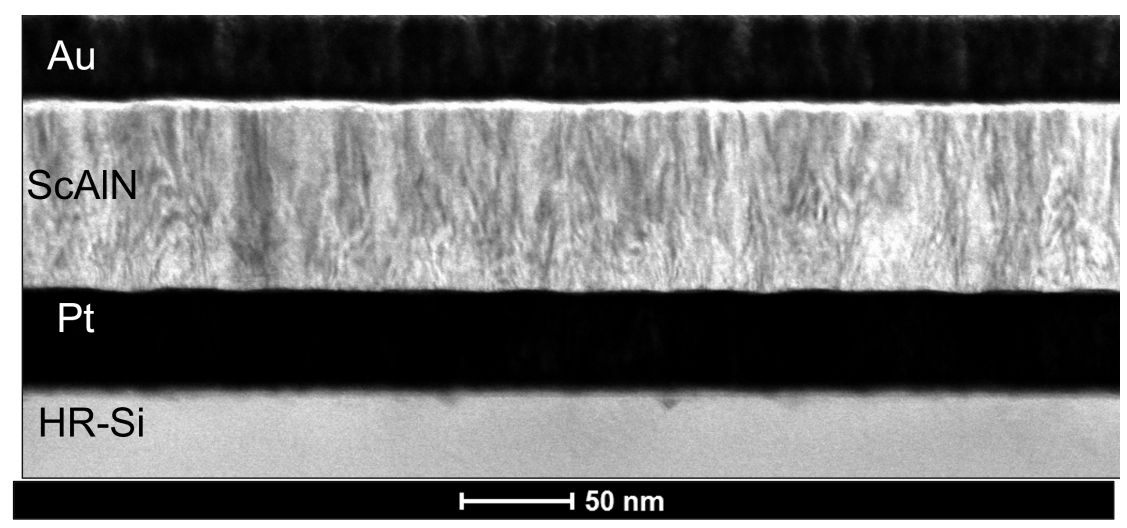

Fig. 8 Cross-sectional TEM of ScAlN BAW resonator stack.

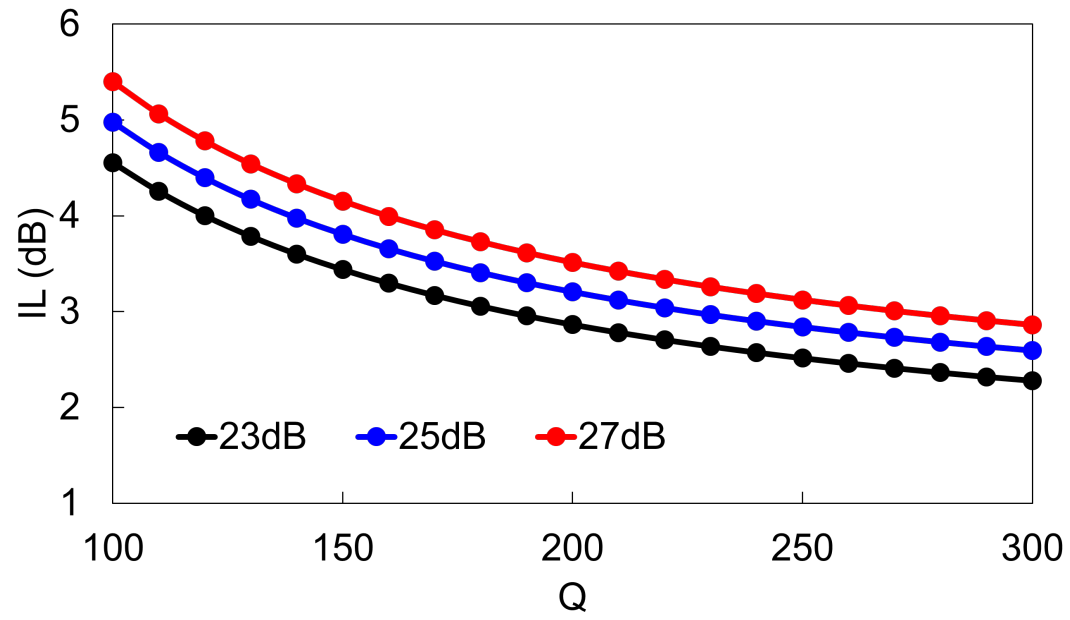

Fig. 9 Achievable IL for different $Q$ in $7^{th}$-order filters of different out-of-band rejection requirements, assuming not impacted by $R_s$.

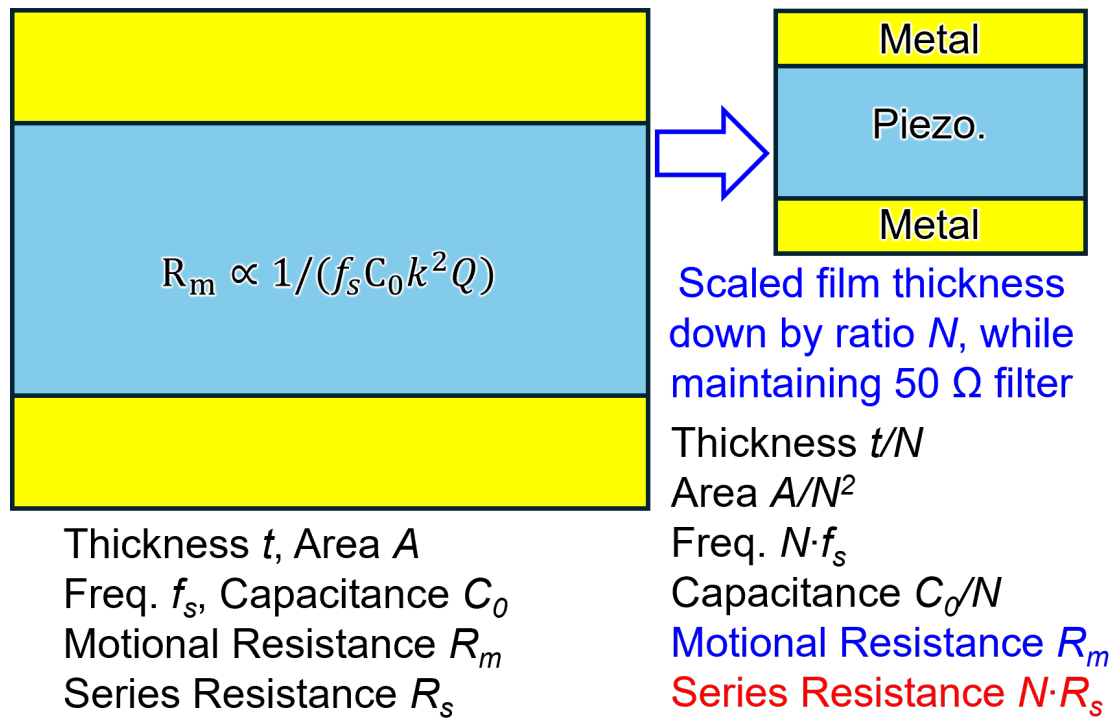

Fig. 10 Change of $R_s$ and $R_m$ according to the piezoelectric film thickness.

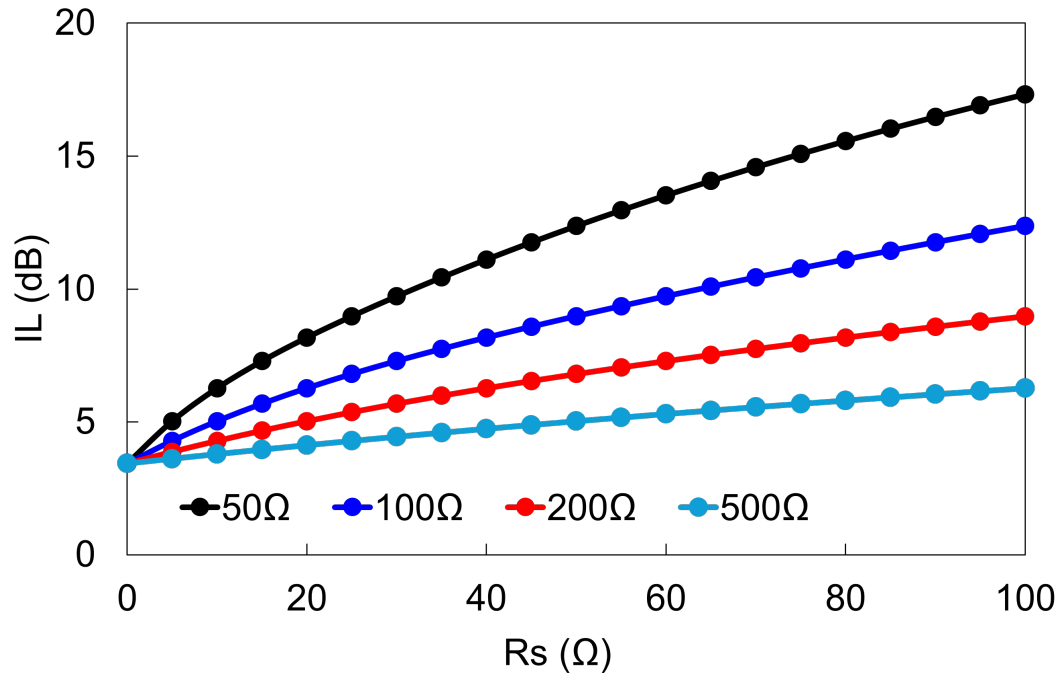

Fig. 11 Achievable IL for different $R_s$ in $7^{th}$-order, 20-dB OoB filters of different port impedance, assuming mechanical $Q$ of 150.

## A. Insertion Loss Decomposition

Currently, the prototype filter is suffering from moderate IL of 6.8 dB. Here, a decomposition of IL is provided. First, the moderate mechanical $Q$ of the resonators, around 150 (see Fig. 6), accounts for most of the loss, predicted to be 3.86 dB (Fig. 3). The impedance mismatch from parasitic capacitance and inductance of the layout, which are not considered in the schematic simulation, leads to an additional insertion loss of 0.6 dB, comparing between Fig. 7 (a) and (b). The remaining 2.34 dB loss is difficult to extract directly, but it is likely due to routing resistance from the thin top electrodes and interconnects. Below, we will provide an analysis of these key factors toward reducing IL in future works.

First, we will discuss the loss contribution from the resonators' moderate mechanical $Q$. Lower $Q$s at higher frequencies are due to the moderate piezoelectric and metal stack film quality. Achieving a high-quality, c-axis-oriented polycrystalline film at these thicknesses is exceptionally difficult [36], [37]. The quality of the sputtered ScAlN film is highly dependent on the seed layer and deposition conditions [38]. As the film becomes thinner, the nucleation phase becomes dominant, and any defects or non-ideal grain orientations in the initial atomic layers have a disproportionately large impact on the film's bulk properties [39]. As shown in the previously reported analysis of sputtered ScAlN below 200 nm, even with an optimal seed layer, achieving perfect texture in ultra-thin films remains a challenge [40]–[43]. A typical full-width at half-maximum (FWHM) is 2.4° [33]. A lower-quality crystal lattice with more grain boundaries and defects, e.g., as shown in a transmission electron microscope (TEM) image of the stack in this work (Fig. 8), leads to increased acoustic scattering (phonon-defect scattering). This acts as an internal friction mechanism, dissipating acoustic energy and fundamentally lowering the material's intrinsic $Q$ [44], [45]. Prior controlled Pt/ScAlN studies [46] show that thinner film thickness degrades crystalline quality (broader FWHM). Because matched thick-film samples are unavailable and chamber conditions have changed, we refer readers to [46] rather than providing a direct SEM/TEM comparison. More SEM/TEM images could be found in related references for ScAlN sputtered on various substrates [47], [48]. To better illustrate the relationship between $Q$ and IL, we plot the achievable IL in a $7^{th}$-order ladder filter for given out-of-band rejection at different mechanical $Q$ in Fig. 9. One can observe that the IL is significantly hurt, especially at high OoB rejection devices. To conclude, this is a primary contributor to the overall low resonator mechanical $Q$ and, consequently, the filter's high IL.

Second, the additional routing resistance $R_s$ of the electrodes is more severe at higher frequencies [49]. The top and bottom electrodes are integral parts of the acoustic cavity and must also be thinned down as the operating frequency increases to enable the high operating frequency [50]. The electrical resistance of a thin film is inversely proportional to its thickness. As the top electrode is scaled down to tens of nanometers, its sheet resistance increases dramatically. This effect is compounded by skin depth and surface scattering effects at high frequencies. The high series resistance of the thin top electrode acts as a parasitic resistor in the electrical path of the resonator [51]. This resistance dissipates a significant portion of the input RF power as heat before it can be efficiently converted into acoustic energy [52]. Fig. 10 indicates the effects on frequency scaling. The ratio of series routing resistance to motional resistance continues to increase as scaling proceeds. This directly degrades the resonator's series resonance Q. Quantitatively, we plot here the impact of $R_s$ on IL for 7th-order 20-dB OoB filter baseline designs with different port impedances (Fig. 11). Filters of different port impedances are plotted to showcase the impact of series resistance. The mechanical $Q$ is assumed as 150. We assume the same $R_s$ on all resonators, which is surely simplified from the real case with distributed routing resistance and different by components. But here is a quick illustration here. It is seen that the 50 Ω devices are susceptible to the electrical loading. Here we estimate an $R_s$ around 10 Ω in our resonators in the prototyping device. While an electrode thickness parametrization could aid intuition, the series resistance at higher frequencies is distributed [49] and geometry-dependent (skin/proximity effects and piezoelectric field distribution), so an accurate thickness to $R_s$ conversion requires layout-specific full-wave EM–piezo co-simulation; accordingly, we report results versus $R_s$ and leave a thickness-based mapping to future work.

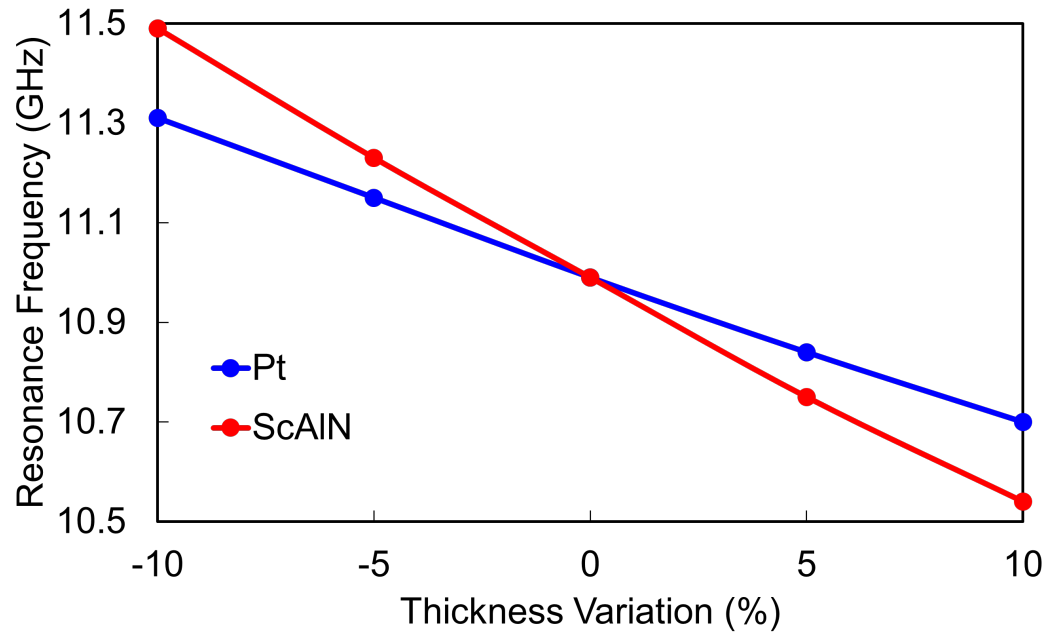

Fig. 12 FEA simulated series resonance of the stack with change of thickness in Pt and ScAlN, using the baseline design in Fig. 2.

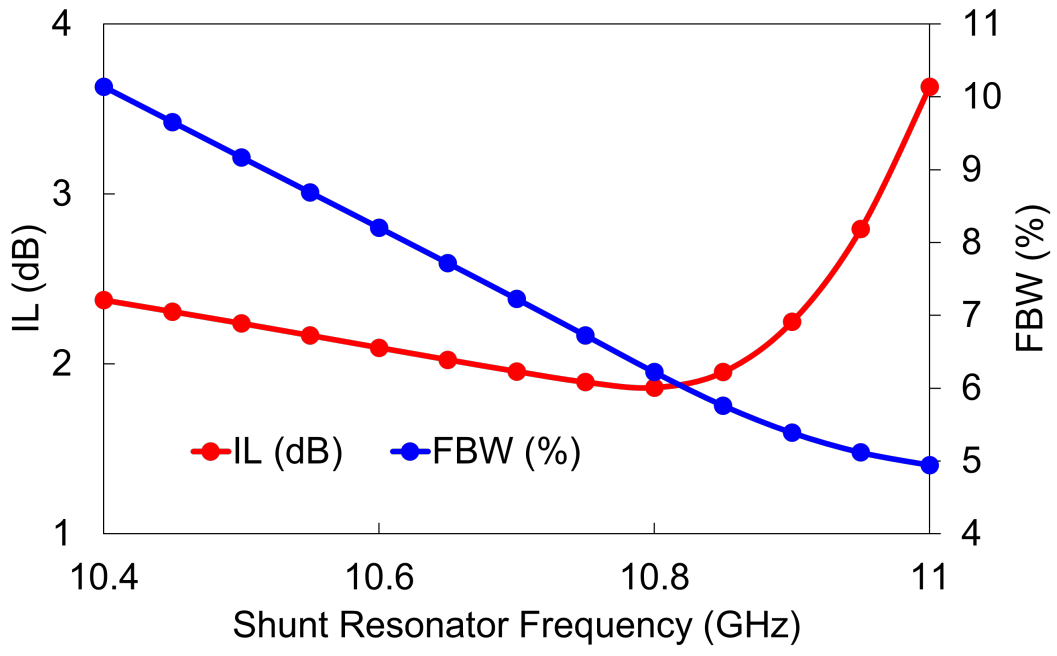

Fig. 13 Impact of IL and BW when the shunt resonator frequency shift deviates from the designed value.

Additional loss due to impedance mismatch could be attributed to layout effects not considered in the schematic layout. Recent research has extensively reported on such effects. EM-acoustic co-simulation will be a useful tool for evaluating and further mitigating the effects [53], [54].

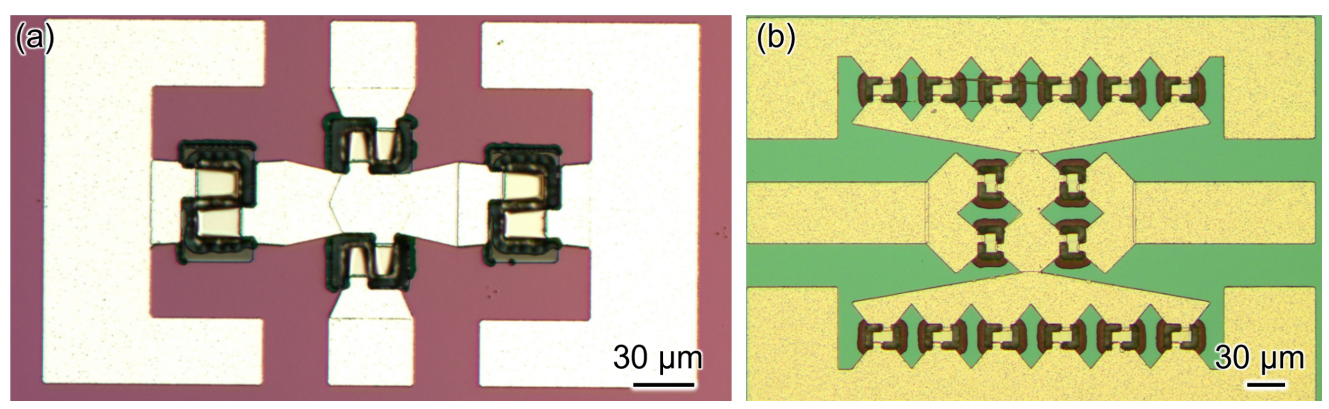

Fig. 14 (a) Collapsed BAW filter by high residual stress on ScAlN. (b) BAW filter with ScAlN film stress management by dividing a big-sized resonator into small-sized resonators connected in parallel.

*B. Center Frequency and Bandwidth Sensitivity*

In addition to the IL, ultra-thin film stacks also require more stringent stack thicknesses to set the center frequency and bandwidth of these filters accurately.

First, the required accuracy of the deposited film thickness is higher. Using the baseline resonator with 85 nm ScAlN sandwiched between 38 nm Pt electrodes, shown in Fig. 2, we plot the impact of thickness variations in different layers on the series resonance in Fig. 12. COMSOL FEA is used for the study. It is seen that the resonance is highly impacted by the stack thickness. What's more, considering that the mechanical properties, e.g., stiffness constants, of such thin piezoelectric and electrodes might vary from those reported for bulk materials, iterations of design and fabrication will be required. Here in the prototype, we are only off the designed center frequency by 0.25 GHz, indicating feasibility of the frequency process upon further improvement.

Second, considering filters are composed of resonators with different resonances, achieving the expected frequency offset is of paramount importance, and also more challenging for the fabrication side than building standalone resonators. If deviated from this frequency offset, the filter BW and IL will be significantly affected. Here, we use the same baseline filter design shown earlier to illustrate the effect. If we fix the series resonator frequency while shifting the shunt resonator resonance, the impact of resonance detuning on both BW and IL can be seen (Fig. 13). As well discussed in prior ladder filter works, too large a frequency shift leads to over coupling, causing a dip in the transmission passband. Fig. 13 should be interpreted together with Fig. 12, which links Pt and ScAlN thickness variations to the corresponding resonance shifts in the shunt resonator; the resonance frequency is highly sensitive as the overall stack thickness is reduced. Given this sensitivity, controlling the loading thickness is essential for accurately setting the intended resonance detuning, which directly governs the achievable FBW.

*C. Stress-Limited Aperture and Impedance Implications*

Additionally, managing the residual stress in sputtered thin films, such as Pt and ScAlN, is a critical factor for device fabrication [55], [56]. While this stress can often be reliably tuned, especially for films thicker than several hundred nanometers, maintaining a tight control range becomes increasingly difficult as the thickness is scaled down [57]. The residual stress within the thin-film stack is then released. If the lateral dimensions of the membrane are too large, the cumulative force from this stress will cause the structure to buckle, wrinkle, or fracture, leading to device failure. An example is shown in Fig. 14 (a)(b), where larger and smaller resonators were fabricated with the same fabrication procedure, but the larger resonators collapse during release, as the release distance is larger. This practical constraint forces designers to use smaller active areas for FBARs. Smaller resonators inherently have higher electrical impedance, which can cause impedance-mismatch losses with the standard 50 Ω system unless connected in parallel. Also, it has been found that a smaller active area leads to lower *Q* for FBARs [58], [59]. Furthermore, smaller devices are more susceptible to parasitic capacitances and resistances, which can further degrade performance and complicate the design of low-loss filters. A dedicated stress study will be needed to explain better the origin and impact of the film thickness [60] but here we focus on reminding readers of one potential caveat in the direct film thickness scaling. Future quantitative studies are needed on the topic.

## IV. ScAlN Filter Frequency Scaling Outlook

The above decomposition isolates the dominant penalties of direct thickness scaling and indicates where effort buys the most IL reduction and frequency control. We therefore conclude with an outlook on materials, electrodes, and topology. These results demonstrate that pushing ScAlN filters toward Ku-band and eventually deeper into the mmWave spectrum with low loss requires more than just dimensional scaling [21], [61]. Future research must focus on a multi-disciplinary approach, targeting fundamental improvements in materials science to enhance thin-film deposition quality (e.g., using molecular beam epitaxy, MBE and metal–organic chemical vapor deposition, MOCVD), developing novel electrode materials with low resistivity at nanoscale thicknesses, e.g., iridium (Ir), ruthenium (Ru), and titanium (Ti), managing film stress, adopting frame resonator design for energy confinement and minimizing loss with recessed and raised frame, minimizing energy leakage into the anchors and substrate [62]–[69]. Efficient stack topologies optimized for overmoding acoustic wave operation, e.g., P3F stacks, can also be alternative solutions in thicker total film stacks with similar $k^2$ [70]–[75]. Addressing these challenges is essential to unlocking the full potential of ScAlN for next-generation wireless systems [21], [23], [33], [61], [76], [77].

## V. Conclusion

We presented an 11.7-GHz, 50-Ω single-layer ScAlN FBAR ladder filter as a quantitative case study of direct thickness scaling, achieving 4.0% FBW and 23.1 dB out-of-band rejection with a minimum IL of 6.8 dB, thereby establishing a realistic performance baseline for this stack and topology. Through measurements and a loss/sensitivity analysis, we attribute the moderate IL and frequency offset primarily to crystal-quality degradation in sub-100-nm ScAlN, series routing resistance in ultrathin electrodes, and stress-limited aperture that constrains impedance and *Q*. This mapping from fabrication-level constraints to device-level metrics (IL, FBW, and center-frequency shift) clarifies where effort is most impactful. Looking forward, improved thin-film deposition, lower-resistivity or thicker effective electrodes with optimized routing, stress-managed and possibly parallelized apertures, and the selective use of heterostructures, e.g., P3F stacks, offer practical paths to reduce loss and tighten frequency control in future mmWave acoustic filters.

## Acknowledgment

The authors thank the DARPA COFFEE program for funding support and Dr. Ben Griffin, Dr. Todd Bauer, and Dr. Zachary Fishman for helpful discussions.